\documentclass[iop]{emulateapj}
\usepackage{color}
\usepackage{savesym}
\savesymbol{tablenum}
\usepackage{amsmath,amssymb,siunitx,ulem,hyperref}
\restoresymbol{aug}{tablenum}
\newcommand{\prx}{Phys. Rev. X}

\shorttitle{Revisiting the lower bound on tidal deformability derived by
AT 2017gfo}
\shortauthors{Kiuchi, Kyutoku, Shibata \& Taniguchi}

\begin{document}

\title{Revisiting the lower bound on tidal deformability derived by AT
2017gfo}
\author{
Kenta Kiuchi\altaffilmark{1,2},
Koutarou Kyutoku\altaffilmark{3,4,5,2},
Masaru Shibata\altaffilmark{1,2},
Keisuke Taniguchi\altaffilmark{6}
}
\altaffiltext{1}{Max Planck Institute for Gravitational Physics (Albert
Einstein Institute), Am M{\"u}hlenberg 1, Potsdam-Golm D-14476, Germany}
\altaffiltext{2}{Center for Gravitational Physics, Yukawa Institute for
Theoretical Physics, Kyoto University, Kyoto, 606-8502, Japan}
\altaffiltext{3}{Theory Center, Institute of Particles and Nuclear
Studies, KEK, Tsukuba 305-0801, Japan}
\altaffiltext{4}{Department of Particle and Nuclear Physics, the
Graduate University for Advanced Studies (Sokendai), Tsukuba 305-0801,
Japan}
\altaffiltext{5}{Interdisciplinary Theoretical and Mathematical Sciences
Program (iTHEMS), RIKEN, Wako, Saitama 351-0198, Japan}
\altaffiltext{6}{Department of Physics, University of the Ryukyus,
Nishihara, Okinawa 903-0213, Japan}

\begin{abstract}
 We revisit the lower bound on binary tidal deformability
 $\tilde{\Lambda}$ imposed by a luminous kilonova/macronova, AT 2017gfo,
 by numerical-relativity simulations of models that are consistent with
 gravitational waves from the binary neutron star merger
 GW170817. Contrary to the claim made in the literature, we find that
 binaries with $\tilde{\Lambda} \lesssim 400$ can explain the luminosity
 of AT 2017gfo, as long as moderate mass ejection from the remnant is
 assumed as had been done in previous work. The reason is that the
 maximum mass of a neutron star is not strongly correlated with the
 tidal deformability of neutron stars with a typical mass of $\approx
 1.4 M_\odot$. If the maximum mass is so large that the binary does not
 collapse into a black hole immediately after merger, the mass of the
 ejecta can be sufficiently large irrespective of the binary tidal
 deformability. We present models of binary mergers with
 $\tilde{\Lambda}$ down to $242$ that satisfy the requirement on the
 mass of the ejecta from the luminosity of AT 2017gfo. We further find
 that the luminosity of AT 2017gfo could be explained by models that do
 not experience bounce after merger. We conclude that the luminosity of
 AT 2017gfo is not very useful for constraining the binary tidal
 deformability. Accurate estimation of the mass ratio will be necessary
 to establish a lower bound using electromagnetic counterparts in the
 future. We also caution that merger simulations that employ a limited
 class of tabulated equations of state could be severely biased due to
 the lack of generality.
\end{abstract}

\maketitle

\section{Introduction} \label{sec:intro}

The first binary neutron star merger was observed as the multi-messenger
event GW170817/GRB 170817A/AT 2017gfo
\citep{ligovirgo2017-3,ligovirgogamma2017,ligovirgoem2017}. Gravitational
and electromagnetic signals have been combined to derive various
information about physics and astrophysics. Examples include the
velocity of gravitational waves \citep{ligovirgogamma2017}, Hubble's
constant \citep{ligovirgoem2017-2}, the central engine of a type of
short gamma-ray burst \citep{mooley_dgnhbfhch2018}, and the origin of
(at least a part of) \textit{r}-process elements
\citep{kasen_mbqr2017,tanaka_etal2017}.

The multi-messenger observations also constrain properties of neutron
stars. Gravitational waves, GW170817, constrain the so-called binary
tidal deformability to $100 \lesssim \tilde{\Lambda} \lesssim 800$,
where precise values depend on the method of analysis and adopted
theoretical waveforms
\citep{de_flbbb2018,ligovirgo2018,ligovirgo2019}. At the same time, some
researchers have argued that the maximum mass of a neutron star
$M_\mathrm{max}$ cannot be significantly larger than $\approx
2.15$--$2.2M_\odot$ based on the electromagnetic features, e.g. the
absence of magnetar-powered radiation
\citep{margalit_metzger2017,shibata_fhkkst2017,rezzolla_mw2018,ruiz_st2018}. \citet{bauswein_jjs2017}
also proposed lower bounds on the radii of massive neutron stars,
assuming that the electromagnetic signals may imply the avoidance of the
prompt collapse. These inferences suggest that supranuclear-density
matter is unlikely to be very stiff.

\citet{radice_pzb2018} proposed a novel idea: $\tilde{\Lambda} \gtrsim
400$ is required to eject material heavier than $0.05 M_\odot$, which
the authors assumed to be required by the high luminosity of AT
2017gfo.\footnote{More precisely, this threshold is derived by fitting
the multi-color evolution of AT 2017gfo.} The logic is that no binary
model with $\tilde{\Lambda} \lesssim 400$ is capable of ejecting $0.05
M_\odot$, even if all of the baryonic remnant can be ejected, in their
numerical-relativity simulations performed with four tabulated equations
of state derived by mean-field theory. This constraint approximately
indicates that neutron stars must be larger than \SI{12}{\kilo\meter}
\citep{zhao_lattimer2018}, and thus it could reject mildly soft
equations of state if reliable. Indeed, this constraint has been used to
infer properties of nuclear matter by various researchers
\citep{most_wrs2018,lim_holt2018,malik_afpajkp2018,burgio_dpsw2018}. Later,
\citet{radice_dai2019} loosened the limit to $\tilde{\Lambda} \gtrsim
300$ by Bayesian inferences; they allowed a standard deviation of 50\%
in the fitting formula of disk masses, which they required to be $> 0.04
M_\odot$, derived using results of
\citet{radice_pzb2018}. \citet{coughlin_dmm2018} also derived a lower
limit of $\tilde{\Lambda} \gtrsim 279$ by Bayesian inferences, with the
improvement of the fit of disk masses via incorporation of the ratio of
the total mass to the threshold mass for the prompt collapse as an
additional parameter. Note that these two works also use other signals,
such as gravitational waves, in a different manner.

\citet{tews_mr2018} critically examined this idea by using
parameterized, general nuclear-matter equations of state. Their key
finding is that the maximum mass is correlated only very weakly with
binary tidal deformability for the masses consistent with GW170817. They
found that some equations of state can support a neutron star with $>2.6
M_\odot$ even if $\tilde{\Lambda}$ is significantly lower than
400. Because the remnant massive neutron star should survive for a long
time, or possibly permanently, after merger for these cases
\citep{hotokezaka_kosk2011,hotokezaka_kkmsst2013}, the argument of
\citet{radice_pzb2018} based on the mass of the ejecta cannot reject
such equations of state and then binary tidal deformability. However,
the maximum mass of a neutron star might also be constrained to
$\lesssim 2.2 M_\odot$ as described above. Whether this constraint on
the maximum mass is compatible with the luminosity of AT 2017gfo is not
trivial.

In this Letter, we demonstrate that the lower bound on $\tilde{\Lambda}$
is not as significant as what \citet{radice_pzb2018} proposed, even if
the maximum mass is only moderately large, $M_\mathrm{max} \lesssim 2.1
M_\odot$, by a suite of numerical-relativity simulations. Specifically,
we find that various models with $\tilde{\Lambda} < 400$ can eject $0.05
M_\odot$ and can explain the luminosity of AT 2017gfo. The models
include asymmetric binary neutron stars with $\tilde{\Lambda} = 242$,
which may not collapse at least until \SI{20}{\milli\second} after
merger. In addition, we also show that the luminosity of AT 2017gfo
could be explained even if the merger remnant does not experience bounce
after merger, when the binary is asymmetric.

\section{Model and equation of state} \label{sec:model}

\begin{table*}
 \caption{Characteristic Quantities of Equations of State Adopted in
 this Work and Results of Simulations} \centering
 \begin{tabular}{cccccccccc} \hline \hline
  $\Gamma$ & $\log P_{14.7}~(\si{dyne.cm^{-2}})$ & $R_{1.35}$ (km) &
  $M_\mathrm{max}~( M_\odot )$ & $q$ & $\tilde{\Lambda}$ & Type &
  $M_\mathrm{dyn}~( M_\odot )$ & $M_\mathrm{disk}~( M_\odot )$ & $\Delta
  x$ (m) \\ \hline
  $3.765$ & $34.1$ & $10.4$ & $2.00$ & $1$ & $208$ & no bounce &
                              $<\num{e-3}$ & $<\num{e-3}$ & $117$ \\
  & & & & $0.774$ & $218$ & no bounce & $<\num{e-3}$ & $0.023$ &
                                      $121$ \\
  $3.887$ & $34.1$ & $10.5$ & $2.05$ & $1$ & $221$ & no bounce &
                              $<\num{e-3}$ & $<\num{e-3}$ & $118$ \\
  & & & & $0.774$ & $230$ & no bounce & $\num{5.2e-3}$ & $0.029$ &
                                      $126$ \\
  $4.007$ & $34.1$ & $10.5$ & $2.10$ & $1$ & $232$ & no bounce &
                              $\num{1.9e-3}$ & $\num{2.7e-3}$ & $118$ \\
  & & & & $0.774$ & $242$ & long & $0.013$ & $0.26\;(0.16,0.097)$ &
                                      $121$ \\
  $3.446$ & $34.2$ & $10.6$ & $2.00$ & $1$ & $232$ & no bounce &
                              $<\num{e-3}$ & $<\num{e-3}$ & $121$ \\
  & & & & $0.774$ & $245$ & no bounce & $\num{2.3e-3}$ & $0.036$ &
                                      $124$ \\
  $3.568$ & $34.2$ & $10.7$ & $2.05$ & $1$ & $247$ & no bounce &
                              $<\num{e-3}$ & $<\num{e-3}$ & $122$ \\
  & & & & $0.774$ & $259$ & no bounce & $0.014$ & $0.038$ & $126$ \\
  $3.687$ & $34.2$ & $10.8$ & $2.10$ & $1$ & $260$ & short &
                              $\num{1.4e-3}$ & $\num{7.8e-3}$ & $124$ \\
  & & & & $0.774$ & $272$ & long & $0.011$ & $0.26\;(0.17,0.092)$ &
                                      $126$ \\
  $3.132$ & $34.3$ & $11.0$ & $2.00$ & $1$ & $272$ & no bounce &
                              $<\num{e-3}$ & $<\num{e-3}$ & $126$\\
  & & & & $0.774$ & $290$ & no bounce & $0.012$ & $0.063$ & $131$ \\
  $3.252$ & $34.3$ & $11.1$ & $2.05$ & $1$ & $288$ & no bounce &
                              $\num{1.2e-3}$ & $\num{1.9e-3}$ & $128$ \\
  & & & & $0.774$ & $305$ & short & $0.015$ & $0.12$ & $131$ \\
  $3.370$ & $34.3$ & $11.1$ & $2.10$ & $1$ & $303$ & short &
                              $\num{2.0e-3}$ & $0.031$ & $128$ \\
  & & & & $0.774$ & $319$ & long & $0.011$ & $0.25\;(0.19,0.12)$ &
                                      $131$ \\
  $2.825$ & $34.4$ & $11.6$ & $2.00$ & $1$ & $345$ & short &
                              $\num{6.5e-3}$ & $0.018$ & $134$ \\
  & & & & $0.774$ & $373$ & short & $0.011$ & $0.087$ & $141$ \\
  $2.942$ & $34.4$ & $11.6$ & $2.05$ & $1$ & $362$ & short &
                              $\num{2.5e-3}$ & $0.016$ & $134$ \\
  & & & & $0.774$ & $387$ & short & $0.011$ & $0.12$ & $139$ \\
  $3.058$ & $34.4$ & $11.6$ & $2.10$ & $1$ & $377$ & long &
                              $\num{9.7e-3}$ & $0.17\;(0.13,0.11)$ &
                                      $134$ \\
  & & & & $0.774$ & $400$ & short & $\num{9.0e-3}$ & $0.16$ & $139$ \\
  $2.528$ & $34.5$ & $12.5$ & $2.00$ & $1$ & $508$ & short &
                              $\num{9.4e-3}$ & $0.053$ & $149$ \\
  & & & & $0.774$ & $558$ & short & $\num{5.6e-3}$ & $0.16$ & $156$ \\
  $2.640$ & $34.5$ & $12.4$ & $2.05$ & $1$ & $516$ & short & $0.012$ &
                                  $0.12$ & $147$ \\
  & & & & $0.774$ & $560$ & short & $\num{6.4e-3}$ & $0.18$ & $154$ \\
  \hline
 \end{tabular}
 \tablecomments{The first and second columns show the parameters that
 specify an equation of state, $\Gamma$ and $\log P_{14.7}$,
 respectively. The third and fourth columns show the circumferential
 radius of a $1.35 M_\odot$ neutron star $R_\mathrm{1.35}$ and the
 maximum mass $M_\mathrm{max}$, respectively. The next five columns
 present models of binaries and results of simulations, where the upper
 and lower rows correspond to equal-mass and unequal-mass binaries,
 respectively. The values of the mass ratio $q$ and binary tidal
 deformability $\tilde{\Lambda}$ are given in the fifth and sixth
 columns, respectively. The seventh, eighth, and ninth columns show the
 fate of the remnant, the mass of dynamical ejecta $M_\mathrm{dyn}$, and
 the mass of the bound material outside the black hole or exceeding
 $\SI{e13}{\gram\per\cubic\cm}$ for the long-lived remnant
 $M_\mathrm{disk}$, for our fiducial $\Gamma_\mathrm{th} = 1.8$. For the
 long-lived remnants, we also show $M_\mathrm{disk}$ for the threshold
 density \SI{e12}{\gram\per\cubic\cm} and \SI{e11}{\gram\per\cubic\cm}
 in the parentheses. The fate is classified into the collapse without
 bounce (no bounce), the short-lived remnant (short), and the long-lived
 remnant (long) as defined in the body text. The tenth column shows the
 grid spacing $\Delta x$ in the finest domain.}  \label{table:model}
\end{table*}

We simulate mergers of equal-mass binaries with $1.375 M_\odot$--$1.375
M_\odot$ and unequal-mass binaries with $1.2 M_\odot$--$1.55
M_\odot$. The total mass, $m_0 = 2.75 M_\odot$, and the mass ratios, $q
= 1$ or $0.774$, are consistent with GW170817
\citep{ligovirgo2017-3,ligovirgo2019} and also with observed Galactic
binary neutron stars \citep[e.g.][]{tauris_etal2017,ferdman2018}. This
should be contrasted with \citet{radice_pzb2018}, where many models are
significantly heavier than GW170817, particularly those with
$\tilde{\Lambda} \lesssim 400$, and the mass ratio is restricted to $q >
0.857$. The initial orbital angular velocity $\Omega$ of the binary is
chosen to be $G m_0 \Omega / c^3 \approx 0.025$ with applying
eccentricity reduction \citep{kyutoku_st2014}, where $G$ and $c$ are the
gravitational constant and the speed of light, respectively. The
binaries spend about six orbits before merger.

Equations of state for neutron star matter are varied systematically by
adopting piecewise polytropes with three segments
\citep{read_lof2009}. This choice allows us to investigate more generic
models rather than particular nuclear-theory models, e.g. mean-field
theory. The low-density segment is identical to that adopted in
\citet{hotokezaka_kosk2011}. The middle-density segment is specified by
pressure at \SI{e14.7}{\gram\per\cubic\cm} denoted by $P_{14.7}$ and an
adiabatic index $\Gamma$. This segment is matched to the low-density
part at the density where the pressure equals. The value of
$P_\mathrm{14.7}$ is known to be correlated with the neutron star radius
\citep{lattimer_prakash2001,read_lof2009}, and we choose $\log
P_\mathrm{14.7}~(\si{dyne.cm^{-2}})$ from $\{ 34.1, 34.2, 34.3, 34.4,
34.5 \}$. The value of $\Gamma$ is determined by, in conjunction with
the high-density segment, requiring the maximum mass of neutron stars to
become $2.00 M_\odot$, $2.05 M_\odot$, and $2.10 M_\odot$. The
high-density segment is given by changing the adiabatic index to 2.8 at
\SI{e15}{\gram\per\cubic\cm}.

The first two columns of Table \ref{table:model} list $\Gamma$ and
$P_\mathrm{14.7}$ for 14 equations of state\footnote{We do not adopt $(
\log P_\mathrm{14.7} , M_\mathrm{max} ) = ( 34.5 , 2.1 M_\odot )$,
because it is unnecessary for our purpose.}  adopted in this study. The
radius of a $1.35 M_\odot$ neutron star and the maximum mass are shown
in the third and fourth columns, respectively. We checked that all of
them are causal; i.e. the sound velocity does not exceed $c$, up to the
central density of the spherical maximum mass configuration. Although
our radii are typically smaller than those favored in
\citet{most_wrs2018}, their probability distribution may be affected
significantly by the small number of available equations of state with
small radii \citep{raithel_op2018}.  As shown in \citet{annala_gkv2018},
our models are compatible with current understanding of nuclear physics
and astronomical observations.

Table \ref{table:model} also presents the binary tidal deformability of
our equal-mass and unequal-mass binaries in the sixth column, where the
mass ratio is given in the fifth column. All are consistent with
constraints obtained by GW170817, irrespective of the details of the
analysis
\citep{ligovirgo2017-3,de_flbbb2018,ligovirgo2018,ligovirgo2019}. As
pointed out by \citet{tews_mr2018}, the binary tidal deformability is
not directly correlated with the maximum mass.

\section{Method of simulations} \label{sec:sim}

Numerical simulations are performed in full general relativity with the
SACRA code \citep{yamamoto_st2008,kiuchi_kksst2017}. The
finite-temperature effect is incorporated by an ideal-gas prescription
following \citet{hotokezaka_kosk2011} with the fiducial value of
$\Gamma_\mathrm{th} = 1.8$, which may be appropriate for capturing the
dynamics of remnant neutron stars \citep{bauswein_jo2010}. We also
performed simulations with $\Gamma_\mathrm{th} = 1.5, 1.6$, and $1.7$
for some models with low values of $\tilde{\Lambda}$; the dependence of
our results on $\Gamma_\mathrm{th}$ will be discussed. Because whether
or not the merger remnant collapses into a black hole in a short time
scale is important for this study, detailed physical effects such as
magnetic fields and neutrino transport are neglected. They are known to
play a central role on a longer time scale than durations of our
simulations, which are performed until 10--$\SI{20}{\milli\second}$
after merger \citep{hotokezaka_kkmsst2013}; thus our results should
depend only weakly on these effects. Although we cannot determine the
electron fraction of the ejecta, which is important to derive
nucleosynthetic yields and characteristics of the kilonova/macronova
\citep{wanajo_snkks2014,tanaka_etal2017,kasen_mbqr2017}, it is not
relevant to the purpose of this work.

We classify the fate of merger remnants into three types. If the remnant
collapses into a black hole without experiencing bounce after merger, we
call it a no-bounce collapse. Note that such collapses are denoted by
the prompt collapse in \citet{bauswein_jjs2017}; we avoid this name,
however, taking into account the fact that some asymmetric models
survive longer than the dynamical time scale up to a few
\si{\milli\second} even if they do not experience bounce. If the remnant
evades the no-bounce collapse but still collapses by
$\SI{20}{\milli\second}$ after merger, it is regarded as a short-lived
remnant. This time scale is approximately identical to that adopted in
\citet{radice_pzb2018}. If the remnant massive neutron star does not
collapse in our simulations, it is called a long-lived remnant. These
three types will be denoted by ``no bounce'', ``short'', and ``long'' in
Table \ref{table:model}, respectively.

We derive the baryonic mass of the unbound dynamical ejecta,
$M_\mathrm{dyn}$, and that of the bound material outside the black hole
or exceeding \SI{e13}{\gram\per\cubic\cm} for the long-lived remnant,
$M_\mathrm{disk}$, from the simulations. The threshold density is chosen
after \citet{radice_pzb2018}, and the dependence of our results on this
value will be described later. The ejecta as a whole should consist of
the dynamical ejecta and the late-time outflow from the merger remnant
\citep[e.g.][]{fernandez_metzger2013,metzger_fernandez2014,just_bagj2015,fujibayashi_knss2018}. Because
our simulations do not include magnetic fields or corresponding
viscosity required to launch the outflow, we simply assume that some
fraction of $M_\mathrm{disk}$ will be ejected by such processes
following \citet{radice_pzb2018}. While \citet{radice_pzb2018}
conservatively (for their purpose) adopted 100\% efficiency for the
ejection from the accretion torus, this efficiency is likely to be lower
than 50\%, particularly when the remnant is a black hole, because the
outflow is a result of the accretion.

\begin{table}
 \caption{Dependence of the Fate of the Remnant, $M_\mathrm{dyn}$, and
 $M_\mathrm{disk}$ on the Grid Spacing, $\Delta x$} \centering
 \begin{tabular}{ccccccc} \hline \hline
  $q$ & $\tilde{\Lambda}$ & $\Delta x$ (m)& Type & $M_\mathrm{dyn}~(
  M_\odot )$ & $M_\mathrm{disk}~( M_\odot )$ \\ \hline
  $1$ & $288$ & $128$ & no bounce & $\num{1.2e-3}$ & $\num{1.9e-3}$ \\
  && $148$ & no bounce & $\num{2.1e-3}$ & $\num{4.8e-3}$\\
  && $164$ & short & $\num{6.9e-3}$ & $0.013$ \\
  $1$ & $508$ & $149$ & short & $\num{9.4e-3}$ & $0.053$ \\
  && $172$ & short & $0.011$ & $0.055$\\
  && $191$ & short & $\num{8.5e-3}$ & $0.045$ \\
  $1$ & $516$ & $147$ & short & $0.012$ & $0.12$ \\
  && $170$ & short & $0.013$ & $0.089$\\
  && $189$ & short & $0.012$ & $0.095$ \\
  $0.774$ & $242$ & $121$ & long & $0.013$ & $0.26$ \\
  && $140$ & long & $0.017$ & $0.26$ \\
  && $156$ & long & $0.019$ & $0.25$ \\
  $0.774$ & $259$ & $128$ & no bounce & $0.014$ & $0.038$ \\
  && $148$ & no bounce & $0.014$ & $0.041$ \\
  && $164$ & short & $0.015$ & $0.31$ \\
  $0.774$ & $290$ & $131$ & no bounce & $0.012$ & $0.063$ \\
  && $152$ & no bounce & $0.013$ & $0.063$ \\
  && $169$ & no bounce & $0.014$ & $0.069$ \\
  $0.774$ & $558$ & $156$ & short & $\num{5.6e-3}$ & $0.16$ \\
  && $180$ & short & $\num{4.7e-3}$ & $0.14$ \\
  && $201$ & short & $\num{4.5e-3}$ & $0.16$ \\
  $0.774$ & $560$ & $154$ & short & $\num{6.4e-3}$ & $0.18$ \\
  && $178$ & short & $\num{5.5e-3}$ & $0.19$ \\
  && $198$ & short & $\num{5.4e-3}$ & $0.15$ \\
  \hline
 \end{tabular}
 \tablecomments{We specify the models by $q$ and $\tilde{\Lambda}$ to be
 compared with those shown in Table \ref{table:model}.}
 \label{table:resolution}
\end{table}

Our results depend weakly on grid resolutions as shown in Table
\ref{table:resolution}. By simulating selected models with three
different resolutions, we estimate that the mass of the ejecta has a
relative error of about a factor of two and an absolute error of
$\num{e-3} M_\odot$ for typical cases with hypothetical first-order
convergence. However, the nominal error reaches an order of magnitude
for marginally stable short-lived remnants, because the fate wanders
from the no-bounce collapse to the short-lived remnant. We think that
this is reasonable and inevitable for models near the threshold, and
these errors should be kept in mind when we discuss implications to AT
2017gfo. In the rest of this Letter, we only show the results of
highest-resolution runs, in which the neutron star radius is covered by
$\approx 65$--70 points with the grid spacing at the finest domain shown
in the tenth column of Table \ref{table:model}.

\section{Result} \label{sec:res}

The merger of binary neutron stars results in dynamical mass ejection
and formation of a remnant, a massive neutron star or a black hole,
surrounded by an accretion torus. Because their dynamics and mechanisms
have been thoroughly described in previous publications
\citep[e.g.][]{hotokezaka_kkosst2013,bauswein_gj2013,radice_glror2016},
we do not repeat detailed explanations. The fate of the merger remnant
(seventh column), the mass of the dynamical ejecta (eighth column), and
the mass of the bound material outside the black hole or exceeding
\SI{e13}{\gram\per\cubic\cm} for the long-lived remnant (ninth column)
are presented in Table \ref{table:model}. The mass of the bound
material, $M_\mathrm{disk}$, for a given equation of state is usually
larger for unequal-mass binaries rather than for equal-mass binaries
because of the efficient tidal interaction and angular momentum transfer
during merger. In particular, some of the asymmetric models leave a
baryonic mass of $\gtrsim 0.03 M_\odot$ even for the no-bounce
collapse. This is because the light components are deformed
significantly before merger and the collapses are gradually induced by
the accretion for these models (see
\url{http://www2.yukawa.kyoto-u.ac.jp/~kenta.kiuchi/GWRC/index.html} for
visualization).

\begin{figure}
 \includegraphics[width=0.95\linewidth]{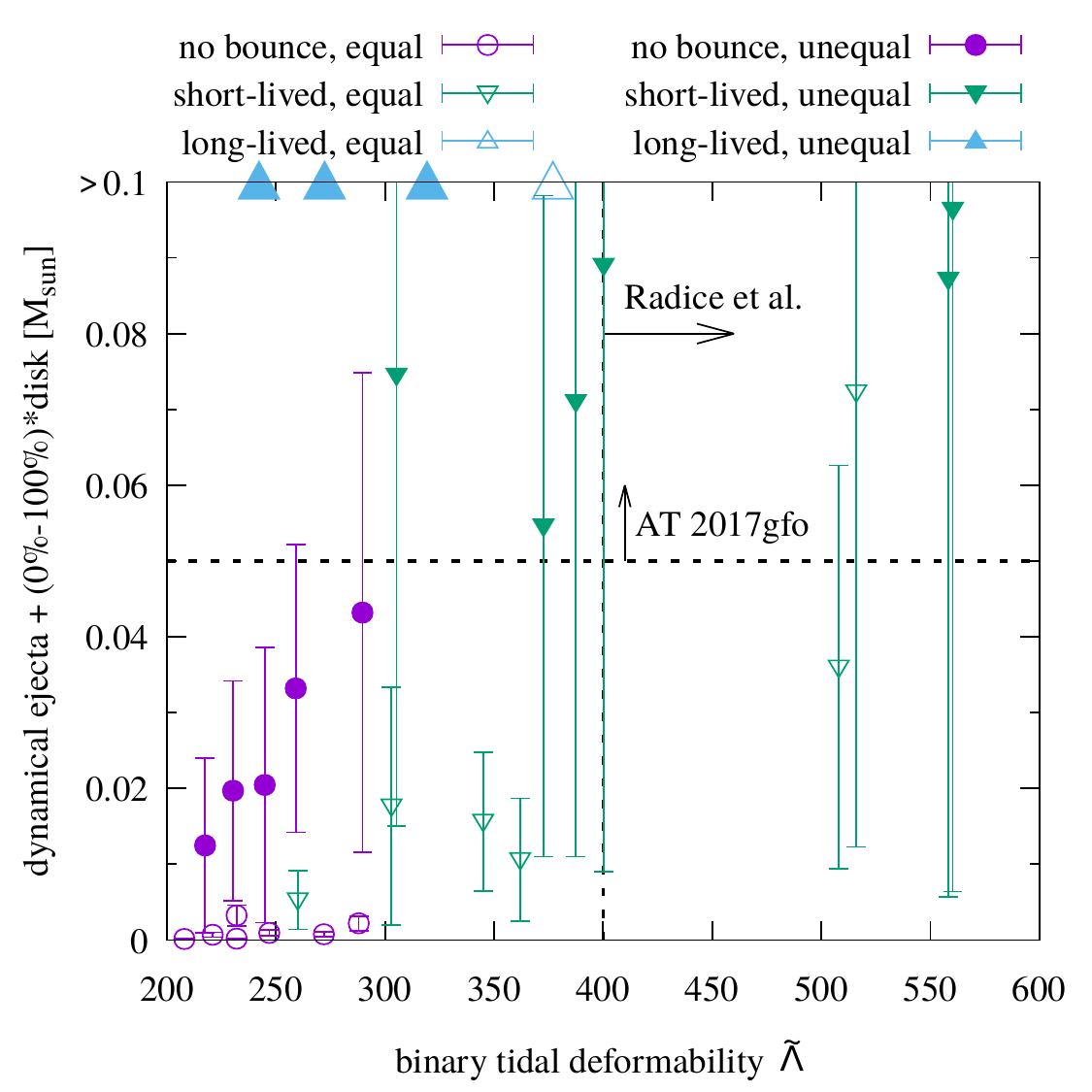} \caption{Mass of the
 ejecta vs. the binary tidal deformability. The errorbars indicate
 ejection of the remnant by from 0\% (i.e. only dynamical mass ejection
 occurs) to 100\% (i.e. all the mass outside the black hole is ejected),
 and the 50\% ejection of the baryonic mass surrounding the black hole
 is marked with symbols. Open and filled symbols denote equal-mass and
 unequal-mass models, respectively. Large triangles on the top axis
 denote the models for which remnant massive neutron stars survive
 longer than \SI{20}{\milli\second} and thus the luminosity of AT
 2017gfo can be explained. Such a model is found even at
 $\tilde{\Lambda} = 242$. The vertical dashed line at $\tilde{\Lambda} =
 400$ is the threshold proposed by \citet{radice_pzb2018}. The
 horizontal dashed line at $0.05 M_\odot$ indicates the mass required to
 explain AT 2017gfo \citep{radice_pzb2018}.} \label{fig:ejecta}
\end{figure}

The masses of the ejecta are summarized visually in
Fig.~\ref{fig:ejecta} against the binary tidal deformability,
$\tilde{\Lambda}$. It is obvious that many binary models with
$\tilde{\Lambda} < 400$ can eject more than $0.05 M_\odot$ and are
capable of explaining the luminosity of AT 2017gfo as far as the mass of
the ejecta is concerned. Indeed, we find that a handful of binary models
with $\tilde{\Lambda} < 400$ result in the formation of a long-lived
remnant, for which $M_\mathrm{disk}$ is always larger than
$0.1M_\odot$. We have verified that the luminosity of AT 2017gfo can be
explained with 50\% ejection efficiency even if the threshold density is
decreased to \SI{e11}{\gram\per\cubic\cm} (see Table
\ref{table:model}). They serve as counterexamples to the claim that
$\tilde{\Lambda} \gtrsim 400$ is required to explain AT 2017gfo
\citep{radice_pzb2018}.

\begin{figure}
 \includegraphics[width=0.95\linewidth]{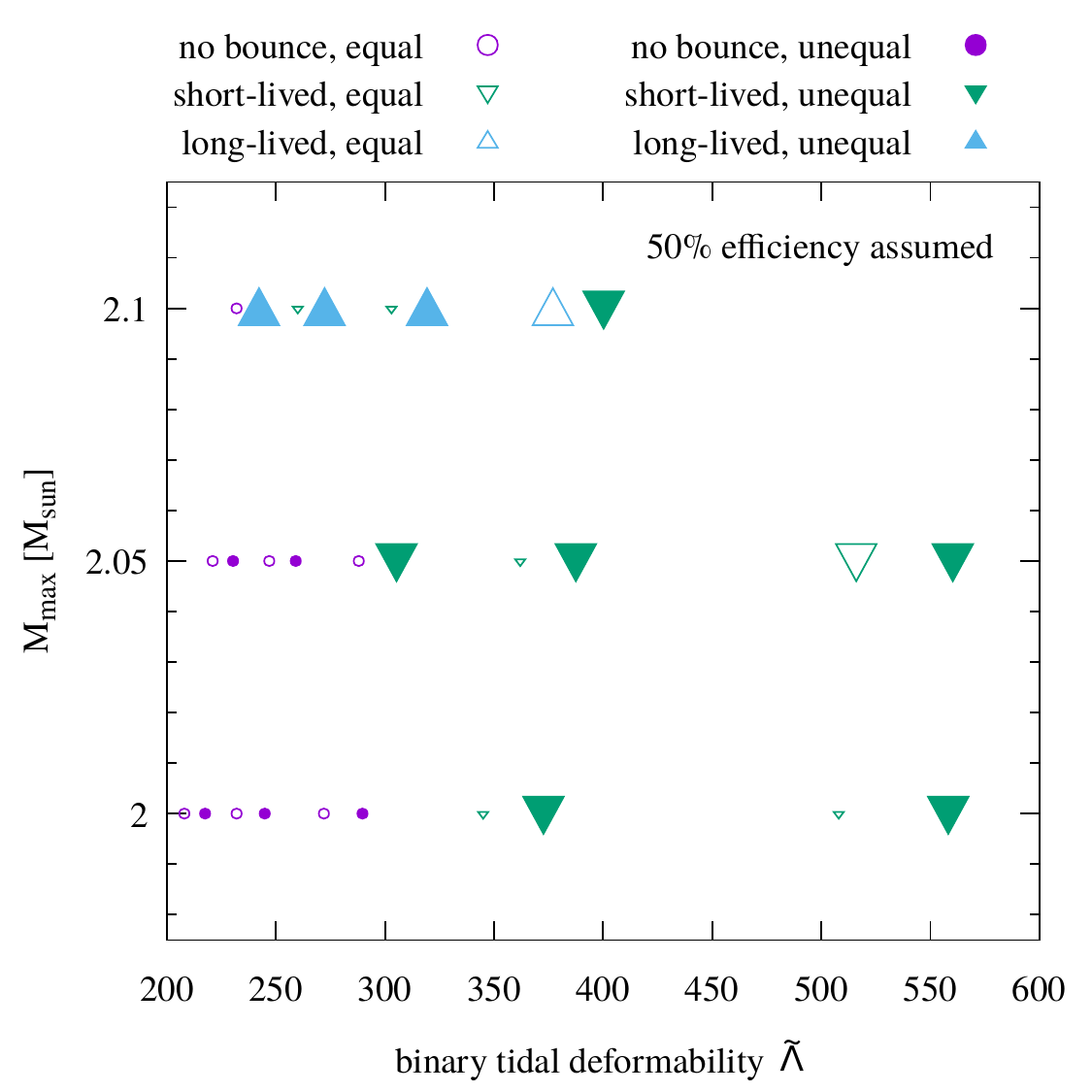} \caption{Summary of
 whether the luminosity of AT 2017gfo can be explained by each model in
 the binary tidal deformability ($\tilde{\Lambda}$)-maximum mass
 ($M_\mathrm{max}$) plane. The large symbols denote models that can
 eject $0.05 M_\odot$ with hypothetical 50\% efficiency and can explain
 the luminosity of AT 2017gfo, and the small ones denote those that
 cannot.} \label{fig:score}
\end{figure}

The key ingredients are the not-so-small maximum mass, $M_\mathrm{max}$,
and the mass asymmetry represented by the small mass ratio, $q$. Their
importance is understood from Fig.~\ref{fig:score}, where we summarize
which model can explain the luminosity of AT 2017gfo in the
$\tilde{\Lambda}$--$M_\mathrm{max}$ plane. Here, we assume a 50\%
ejection efficiency of the bound material for concreteness. On one hand,
for the case that the maximum mass is $2 M_\odot$, all the models
collapse by \SI{20}{\milli\second} after merger. Not only equal-mass
models have no chance of ejecting $0.05 M_\odot$,\footnote{A model with
$\tilde{\Lambda} = 508$ can eject $0.05 M_\odot$ if the efficiency
exceeds 77\%.} but also the mass asymmetry of $q = 0.774$ does not save
any model with $\tilde{\Lambda} < 377$. On the other hand, if the
maximum mass is as large as $2.1 M_\odot$, many models produce
long-lived remnants. Actually, all the asymmetric binaries considered
here are capable of explaining the luminosity of AT 2017gfo. The lowest
value of $\tilde{\Lambda}$ of models that can eject $0.05 M_\odot$ is
242. Figure \ref{fig:score} suggests that, if $M_\mathrm{max}$ is larger
than $2.1 M_\odot$, then the lower bound on $\tilde{\Lambda}$ derived by
AT 2017gfo may become looser than that found in this study.

We also find that all the models with $\tilde{\Lambda} > 400$ are
capable of ejecting $0.05 M_\odot$ if 100\% ejection efficiency is
adopted. This is consistent with the findings of \citet{radice_pzb2018}.

\begin{table}
 \caption{Dependence of the Fate of the Remnant, $M_\mathrm{dyn}$, and
 $M_\mathrm{disk}$ on $\Gamma_\mathrm{th}$} \centering
 \begin{tabular}{ccccccc} \hline
  $q$ & $\tilde{\Lambda}$ & $\Gamma_\mathrm{th}$ & type &
  $M_\mathrm{dyn}~[M_\odot]$ & $M_\mathrm{disk}~[M_\odot]$ \\ \hline
  $0.774$ & $242$ & $1.8$ & long & $0.013$ & $0.26$ \\
  && $1.7$ & short & $0.011$ & $0.045$ \\
  && $1.6$ & short & $\num{7.6e-3}$ & $0.036$ \\
  && $1.5$ & short & $\num{6.5e-3}$ & $0.033$ \\
  $0.774$ & $272$ & $1.8$ & long & $0.011$ & $0.26$ \\
  && $1.7$ & long & $0.013$ & $0.26$ \\
  && $1.6$ & long & $0.014$ & $0.27$ \\
  && $1.5$ & short & $\num{9.8e-3}$ & $0.042$ \\ \hline
 \end{tabular}
 \tablecomments{We specify the models by $q(=0.774)$ and
 $\tilde{\Lambda}$ to be compared with those shown in Table
 \ref{table:model}.} \label{table:thermal}
\end{table}

The fate of the merger remnant depends on the strength of the
finite-temperature effect for marginal cases. For example, the lowest
value of $\tilde{\Lambda}$ that can explain the luminosity of AT 2017gfo
is $242$ in our models if the fiducial $\Gamma_\mathrm{th} = 1.8$ is
adopted, where the outcome is a long-lived remnant. However, the remnant
becomes short lived for $\Gamma_\mathrm{th} \le 1.7$ because of the
reduced thermal pressure and fails to eject $0.05 M_\odot$. This
indicates that the finite-temperature effect must be moderately strong
for this model to account for AT 2017gfo. We also find that the model
with $\tilde{\Lambda} = 272$ results in the long-lived remnant only when
$\Gamma_\mathrm{th} \ge 1.6$, whereas the short-lived remnant for a very
small value of $\Gamma_\mathrm{th} = 1.5$ can eject $0.05 M_\odot$ if
100\% efficiency is assumed. The results for them are summarized in
Table \ref{table:thermal}. Although our conclusion that binaries with
$\tilde{\Lambda} \lesssim 400$ are capable of explaining the luminosity
of AT 2017gfo is unchanged, these observations imply that accurate
incorporation of the finite-temperature effect is also crucial to infer
precise properties of the zero-temperature equation of state from
electromagnetic counterparts.

\section{Discussion} \label{sec:disc}

We conclude that the lower bound on binary tidal deformability is
$\tilde{\Lambda} \le 242$ if an ejection of $0.05 M_\odot$ is
required. We speculate that lower values of $\tilde{\Lambda}$ than this
could even be acceptable if we employ an equation of state that supports
a maximum mass larger than $2.1 M_\odot$ and/or increase the degree of
asymmetry. The precise value of the threshold depends also on the
strength of the finite-temperature effect, represented by
$\Gamma_\mathrm{th}$ in our study.

We also find that an asymmetric binary that results in a no-bounce
collapse can explain the luminosity of AT 2017gfo, if moderately high
$\approx 60\%$ ejection efficiency from the remnant is admitted. The
lower bounds proposed in \citet{bauswein_jjs2017} are satisfied for the
equation of state of this model, with which the radii of $1.6 M_\odot$
and maximum-mass configurations are $10.93$ and \SI{9.66}{\km},
respectively. However, our finding would potentially invalidate the
argument of \citet{bauswein_jjs2017} and its future application.

\begin{figure}
 \includegraphics[width=0.95\linewidth]{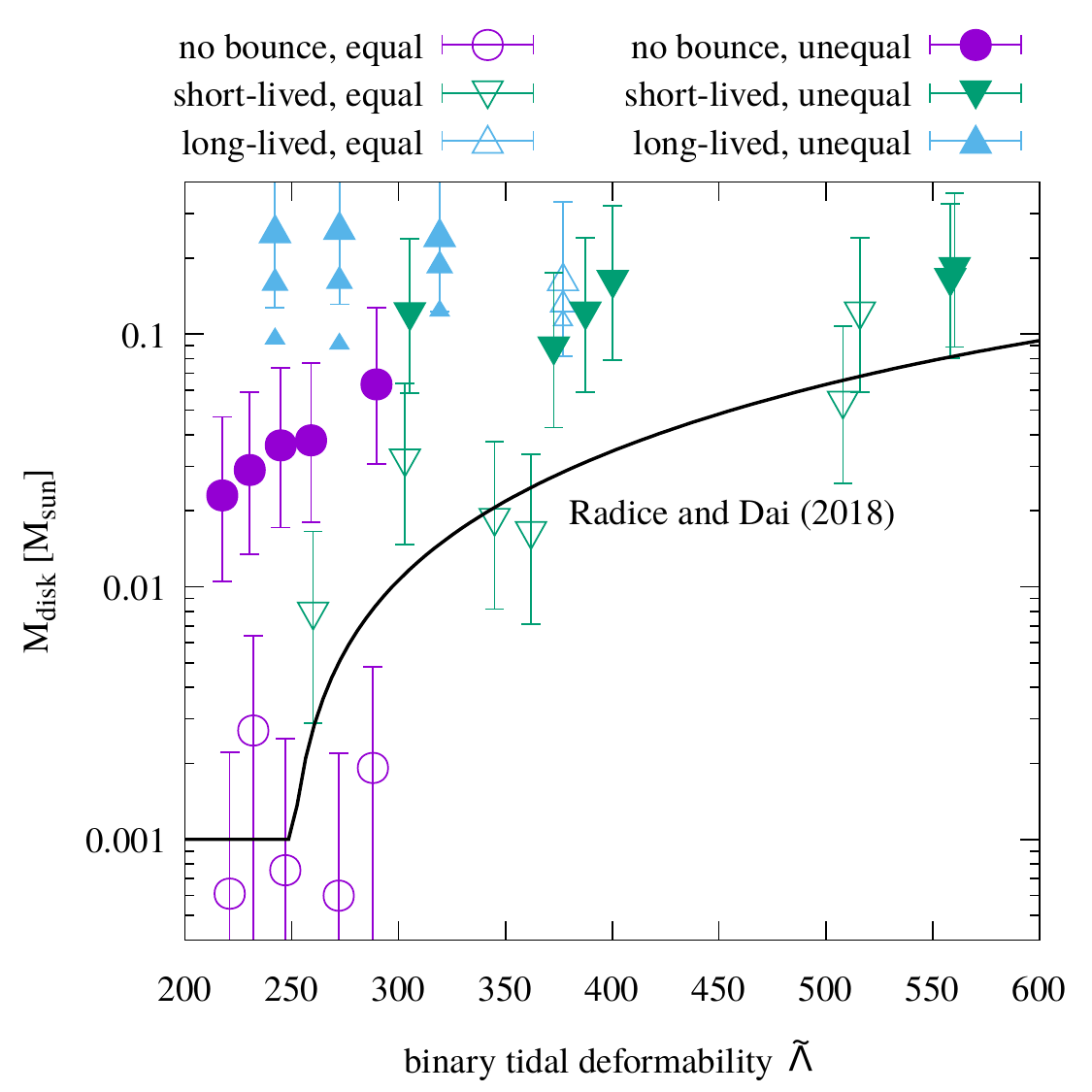} \caption{Disk mass
 vs. the binary tidal deformability. The errorbars denote the typical
 relative error of a factor of two and absolute error of $\num{e-3}
 M_\odot$ (see Sec.~\ref{sec:sim}). The values for the threshold density
 of \SI{e12}{\gram\per\cubic\cm} and \SI{e11}{\gram\per\cubic\cm} are
 shown with small symbols for long-lived remnants. We also show the fit
 derived in \cite{radice_dai2019}. The correlation between
 $M_\mathrm{disk}$ and $\tilde{\Lambda}$ is not significant in our
 models, and the applicability of the fit due to \cite{radice_dai2019}
 is very limited.} \label{fig:disk}
\end{figure}

Our results indicate that the mass ratio is critically important to
derive reliable constraints on neutron star properties from
electromagnetic emission as also argued in \citet{radice_pzb2018}. If
the binary turns out to be symmetric, it is possible that
$\tilde{\Lambda} \gtrsim 400$ is necessary as \cite{radice_pzb2018}
originally proposed. Indeed, we find no symmetric model with
$\tilde{\Lambda} < 377$ that can eject $0.05 M_\odot$. However,
Fig.~\ref{fig:disk} shows that the mass asymmetry significantly obscures
the correlation between the disk mass and binary tidal deformability,
which is the basis of previous attempts to constrain $\tilde{\Lambda}$
from AT 2017gfo. In light of our results, fitting formulas adopted in
\citet{radice_dai2019} and \citet{coughlin_dmm2018} have severe
systematic errors. Further investigation is required to clarify
precisely the effect of asymmetry. Although the mass ratio can be
determined from gravitational-wave data analysis, the degeneracy with
the spin must be resolved to achieve high precision
\citep{hannam_bffh2013}.

The velocity and the composition can potentially be used as additional
information to examine binary models. Some previous work attempted to
associate either the blue or red component of AT 2017gfo to dynamical
ejecta to improve parameter estimation
\citep{gao_caz2017,coughlin_etal2018}. However, the derived binary
parameters, in particular the mass ratio, disagree between these
works. As shown by \citet{kawaguchi_st2018}, such an association is not
necessarily justified once interaction among multiple ejecta components
is taken into account. Detailed modelings of the emission are required
if we would like to utilize the velocity and/or the composition to put
constraints on properties of neutron stars.

Another lesson drawn from our study is that the possible parameter space
of nuclear physics may not be satisfactorily covered by current
tabulated equations of state \citep{tews_mr2018}. For example, equations
of state derived by relativistic mean-field theory tend to predict a
large maximum mass only when the typical radius is large
\citep{radice_phfbr2018}, and thus the value of binary tidal
deformability is also high. Such a correlation is not likely to be
physical but ascribed to the method of quantum many-body
calculations. Specifically, the large maximum mass and the small radius
can be accommodated in variational calculations
\citep[e.g.][]{togashi_ntyst2017}. As Fig.~\ref{fig:score} shows, the
outcome of the merger depends significantly on the maximum mass, even if
the binary tidal deformability is unchanged. It should be remarked that
models with $\tilde{\Lambda} < 400$ of \citet{radice_pzb2018} are
generated by assigning total masses larger than those allowed by
GW170817 \citep{ligovirgo2017-3,ligovirgo2019} except for the SFHo
equation of state \citep{steiner_hf2013}. It is impossible for other
equations of state adopted by them to produce binary models equipped
with $\tilde{\Lambda} \lesssim 400$ and the total mass allowed by
GW170817 simultaneously. This feature artificially enhances the chance
of the early collapse. If we wish to put reliable constraints on neutron
stars via numerical simulations, care must be taken regarding the
limitation of the adopted models including the finite-temperature
effect.

\begin{acknowledgments}
 We thank Andreas Bauswein, Sebastiano Bernuzzi, Kenta Hotokezaka, David
 Radice, and Masaomi Tanaka for valuable comments. Numerical
 computations were performed at Oakforest-PACS at Information Technology
 Center of the University of Tokyo, Cray XC50 at CfCA of National
 Astronomical Observatory of Japan, and Cray XC30 at Yukawa Institute
 for Theoretical Physics of Kyoto University. This work is supported by
 Japanese Society for the Promotion of Science (JSPS) KAKENHI grant Nos.
 JP16H02183, JP16H06342, JP17H01131, JP17K05447, JP17H06361, JP18H01213,
 JP18H04595, and JP18H05236, and by a post-K project hp180179.
\end{acknowledgments}

%\bibliography{ms}

\end{document}